\documentclass[12pt]{article}

\usepackage{amsmath}
\usepackage{amssymb}
\usepackage{amsfonts}
\usepackage{latexsym}
\usepackage{graphicx}
\usepackage{latexsym}

\usepackage{hyperref}
\usepackage{indentfirst} 

\def\ln{\,\mbox{ln}\,}

\def\det{\,\mbox{det}\,}
\def\Det{\,\mbox{Det}\,}
\def\tr{\,\mbox{tr}\,}
\def\Tr{\,\mbox{Tr}\,}

\def\diag{\,\mbox{diag}\,}

\def\al{\alpha}
\def\be{\beta}

\def\Ga{\Gamma}
\def\de{\delta}
\def\De{\Delta}

\def\la{\lambda}
\def\La{\Lambda}
\def\si{\sigma}

\def\ph{\varphi}

\def\om{\omega}

\def\na{\nabla}
\def\pa{\partial}

\def\beq{\begin{eqnarray}}
\def\eeq{\end{eqnarray}}
\newcommand{\eq}[1]{(\ref{#1})}
\newcommand{\n}[1]{\label{#1}}

\begin{document}

\begin{center}

{\Large One-loop renormalization of Lorentz/CPT violating scalar 
field theory in curved space-time}

\vskip 6mm


\textbf{Tib\'{e}rio de Paula Netto}$^a$
\footnote{ E-mail: tiberiop@fisica.ufjf.br}

\vskip 4mm

(a) \ Departamento de F\'{\i}sica, \ ICE, \
Universidade Federal de Juiz de Fora
\\ Juiz de Fora, \ 36036-330, \ MG, \ Brazil.

\end{center}

\vskip 6mm

\begin{quotation}
\noindent
{\large {\it Abstract}}.
\quad
The one-loop divergences for the scalar field theory with Lorentz 
and/or CPT breaking terms are obtained in curved space-time. We 
analyze two separate cases: minimal coupled scalar field with gravity 
and nonminimal one. For the minimal case with real scalar field, the 
counterterms are evaluated in a nonperturbative form in the CPT-even 
parameter through a redefinition of space-time metric. In the most 
complicated case of complex scalar field nonminimally interacting with 
gravity, the solution for the divergences is obtained in the first 
order in the weak Lorentz violating parameter. The necessary form of 
the vacuum counterterms indicate the most important structures of 
Lorentz/CPT violations in the pure gravitational sector of theory. 
The conformal theory limit is also analyzed. It turns out that if we 
allow the violating fields to transform, the classical conformal 
invariance of massless scalar fields can be maintained in the
$\,\xi = 1/6\,$ case. At quantum level the conformal symmetry is 
violated by trace anomaly. As a result, conformal anomaly and the 
anomaly-induced effective action are evaluated in the presence of 
extra Lorentz- and/or CPT-violating parameters. Such gravitational 
effective action is important for cosmological applications and can be 
used for searching of Lorentz violation in primordial universe in the 
cosmological perturbations, especially gravitational waves. 
\end{quotation}

\section{Introduction}
\label{int}

In the last years there was an intensive investigation of
theoretical and experimental aspects in theories where the Lorentz 
and/or CPT symmetries are violated. Such violations could emerge in a 
number of different fashions, most of them related to a new physics 
at the Planck scale $M_P \simeq 10^{19} GeV$. As an example we can 
cite quantum effects in string theory \cite{String} and loop quantum 
gravity \cite{LoopQG}, geometrical effects as noncommutativity 
\cite{noncommutative}, torsion \cite{torsion} and nonmetricity 
\cite{Foster:2016uui}; and so on \cite{else}. Regardless of wherever 
these violations might come from, or if exposed scenarios may or may 
not occur, the experimental/observational searching of remaining small 
deviations from Lorentz and CPT symmetries in nowadays attainable
scales are of crucial importance, representing at the moment a very 
active area of physics \cite{Kostelecky:2016ufw}. Such considerations 
can indicate the existence of new physical phenomena or, at least, 
improve our understanding of the limits of validity of the current one.

The conventional approach for this type of consideration starts, of 
course, with theoretical aspects, considering the most general 
consistent form of Lorentz and CPT violating terms in the action of 
quantum matter fields. Such theory is called the standard model 
extension (SME) \cite{SME}. The action of SME contains all possible 
new operators that parameterize the possible Lorentz and/or CPT 
violations which can be constructed from scalars, fermions and 
vector fields using the effective quantum field theory (QFT) approach.
After that, as a next step one can look for the possible 
phenomenological manifestation of these new terms. For the updated 
date table of bounds on Lorentz/CPT breaking terms and some 
experimental details see \cite{Kostelecky:2008ts} and further 
references therein. Likewise, from the QFT perspective, the presence 
of additional background fields means that the modifications may occur 
not only at classical level but also at the quantum one. The 
investigation of flat space-time loop effects in Lorentz and CPT 
violated quantum electrodynamics (QED) started in the pioneer 
work \cite{Kostelecky:2001jc}, where the corresponding quantum effects 
were derived and new bounds on the violating parameter were
indicated.

From the pure gravitational side there is also the possibility for 
Lorentz violation in the SME \cite{GravityCPTL}. Nonetheless, in the 
vacuum sector of Lorentz and CPT violating theories one can introduce 
terms with these symmetry breaking in many distinct ways. Besides the 
linear in curvature violating terms, there is additionally the 
possibility to introduce a huge amount of higher derivative structures 
in the gravitation action. Indeed, the pure gravitational sector with 
Lorentz violation can be described by an action containing an infinite 
series of higher derivative terms with operators of increasing mass 
dimension. As a concrete example we can cite the theories with 
torsion. In this case, the general vacuum action of gravity, which 
contains only a small part of CPT and Lorentz violating terms, 
includes incredible 168 independent terms \cite{Chris80}. Such great 
arbitrariness makes it very difficult to choose which of those terms 
are the most important ones and, therefore, a natural question is how 
to define the form of the possible Lorentz and/or CPT symmetries 
violation in the gravitational sector. 

One of the possibilities is to introduce only those terms which can 
emerge as vacuum divergences in a semiclassical theory of quantum 
matter fields. These criteria concern the minimal necessary set 
of terms in the classical gravitational action which are required by 
renormalizability. One example where this approach is widely applied 
is the QFT in curved space-time. In this case the renormalizability is 
achieved by introducing along with the Einstein-Hilbert and 
cosmological constant terms, a full set of local covariant fourth 
derivative structures (for an introduction on this subject see, e.g., 
the textbook \cite{book}). For the theories with Lorentz- and/or 
CPT-violating terms one meets an analogous situation. The introduction 
of new terms in the matter fields sector requires, at the quantum 
level, the extension of classical action of vacuum with a set of 
structures depending on the violating parameters. The form of those 
terms may be established on the basis of general covariance and power 
counting arguments, but only direct calculation of the counterterms can 
indicate which terms are truly necessary. And since the Lorentz and 
CPT breaking terms are very small, the one-loop calculation are the 
most important ones. Hence, our strategy to investigate the possible 
role of the violating parameters in the gravitational sector is to start 
by deriving the one-loop divergences for the SME fields on curved 
background. 

The first consideration in this direction was presented in 
Ref.~\cite{CPTLGuiSh} (see also \cite{CPTLSh}). In this work the 
one-loop calculations in the Lorentz and CPT violating QED were 
considered in curved space-time. However, the presented calculations 
were not complete, because only the divergences of effective 
action (EA) which concern to the minimal part of the corresponding 
bilinear operator have been taken into account. After that, the problem 
of working with the complicated nonminimal structures which appears 
typically in the EA of Lorentz and CPT violating theories was solved 
in Ref.~\cite{CPTLTiSh}, by introducing a new calculation trick 
involving the inversion of the minimal part of bilinear operator. 
Working at first order in the violating parameters, the complete 
photon contribution to vacuum renormalization was obtained. In this 
work the corresponding EA of gravity was also derived by integration 
of conformal anomaly. This anomaly-induced EA corresponds to the 
leading quantum contribution for present-day low energy physics 
applications, since the photon is the lightest field, and the other 
massive fields suffer from the Appelquist and Carazzone decoupling 
theorem \cite{Appelquist:1974tg}, which takes place, also, for the 
vacuum gravitational sector \cite{Gorbar:2002pw}. However, for the 
interesting cosmological applications in the early universe, in which 
the typical energies of physical phenomena are very large, all other 
quantum matter fields provides quantum contributions which are as 
important as the photon, since in this high energy situation matter 
behaves approximately as free radiation. Consequently, the evaluation 
of loop effects in curved background coming from the other SME sectors 
is also a relevant question.

In the present work we report the results of the one-loop counterterms 
calculations for the Lorentz and CPT violating massive complex scalar 
field theory in curved space-time. The effects in the vacuum 
renormalization of the adimensional CPT-even violating field and mass 
dimension CPT-odd parameter are analyzed. Furthermore, we also 
consider the possibility of nonminimal interaction of scalars with 
gravity in the form $\,\xi R\ph^* \ph\,$. The inclusion of nonminimal 
term is necessary for the renormalizability of an interacting theory 
which includes scalars in curved background without Lorentz and/or 
CPT-violating terms (see, e.g., \cite{book,birdav} for the 
introduction). Furthermore, the nonminimal parameter $\xi$ plays an 
important role to inflationary models such as Higgs inflation, 
where the nonminimal $\xi R H^\dagger H $-term is added to the Higgs 
potential \cite{Higgs1,Higgs2}. It seems natural to extend these theories to the quantum level, 
studying the possible 
interaction between the nonminimal parameter and the Lorentz/CPT 
violating fields, especially because we know that quantum effects 
are essential in the scalar inflaton models. 

The introduction of nonminimal interaction also opens the way to the 
study the massless conformal theory limit and conformal anomaly 
\cite{Duff94,ConfPo}. The integration of the anomaly yields the anomaly-induced effective action of gravity 
\cite{Ano-int}, which is a compact analytic form of quantum 
corrections. The anomaly-induced EA has many useful applications in cosmological models such as the full 
Starobinsky 
model of inflation \cite{star} or its modified version \cite{MSt}. The 
primordial universe could be seen as a subject of very special interest 
for the Lorentz and CPT symmetries violating theories, because it can 
be considered as a laboratory for the study of physical phenomena in 
energy scales not currently available in particle  
accelerators on Earth. Moreover, the early universe may have been very 
different from its present-day situation, because since then some kind of physical process of space-time symmetry restoration may have occurred.
We expected that the Lorentz breaking terms make no effect on the 
zero-order homogeneous and isotropic cosmology, since the violating 
fields define a preferable direction in the space-time. However, many 
of the symmetry-breaking terms may lead to anisotropy in the cosmic 
microwave radiation \cite{Kostelecky:2007fx}, coming from the cosmic 
perturbations in the inflationary epoch. Therefore, it would be 
interesting to evaluate the possibility of such violations, in 
particular, in the primordial universe with gravitational waves 
calculations.

The organization of the paper is as follows. In Sec.~\ref{sec-div} 
the one-loop divergences for Lorentz- and/or CPT-violating scalar 
field theory are derived in curved space-time. We consider separately 
minimal and nonminimal interaction with gravity. In both cases we 
adopt dimensional regularization and the curved background 
calculations are performed by means of the heat kernel techniques 
related with the Schwinger-DeWitt method \cite{dewitt,bavi85}. Hence, 
the minimal set of pure gravitational terms requested by 
renormalizability is also analyzed. In Sec.~\ref{sec-CS} the local 
conformal symmetry limit is investigated in the presence of the 
symmetry-breaking terms, and from the results obtained in the previous 
section, the conformal anomaly is calculated. Sec.~\ref{sec-aIEA} is 
devoted to integrating conformal anomaly and, therefore, the 
gravitational anomaly-induced effective action is derived. From the 
technical side most of the consideration in this section are pretty 
well known and the standard procedure do not change so much in the 
presence of Lorentz violating fields, but we present adequate details 
in order to make it readable for those not trained within this 
subject. Finally, in Sec.~\ref{con} we draw our conclusions.

Our sign conventions are $\,\eta_{\mu\nu} \,=\, \diag(+,-,-,-)\,$ for 
Minkowski space-time metric and 
$\,R^{\al}_{\,.\,\be\mu\nu} \,=\, \pa_\mu \Ga^{\al}_{\be\nu} - \dots\,$
for the Riemann tensor. The Ricci tensor is 
$\,R_{\mu\nu} \,=\,R^{\al}_{\,.\,\mu\al\nu}\,$
and $R \,=\, g^{\mu\nu} R_{\mu\nu} $ denotes the Ricci scalar curvature.
We also assume that the 
space-time is torsionless and use spatial distance and mass 
definitions such that $c\,=\,\hbar\,=\,1$.

\section{Derivation of one-loop divergences}
\label{sec-div}

Our model of interest is the massive complex scalar field theory with 
extra Lorentz and CPT symmetry-breaking terms. The extension for 
curved background is obtained by nonminimal procedure of covariant 
generalization. The corresponding action has the form
\beq
\n{action}
S = \int d^4x \, \sqrt{-g} \, \left\{ 
g^{\mu\nu} \pa_\mu \ph^* \, \pa_\nu \ph
- \, m^2 \ph^* \ph
+ \xi  R\, \ph^* \ph
+  K^{\mu\nu}(x) \, \pa_\mu \ph^* \, \pa_\nu \ph
+ k^\mu (x) \, j_\mu 
 \right\},
\eeq
where 
\beq
j_\mu  \,=\, i\, \left( 
\ph^* \, \pa_\mu \ph - \ph \, \pa_\mu \ph^* 
\right)
\eeq 
is the scalar field current. The $\,\xi R\ph^* \ph\,$-term is 
called nonminimal and the dimensionless parameter  $\xi$ is known as 
nonminimal parameter. The dimensionless $K^{\mu\nu}(x)$-term is the 
Lorentz CPT-even violating second-rank tensor and $k^\mu (x)$ is the 
Lorentz CPT-odd breaking parameter with mass dimension. Since we are 
working in a curved background, we do not consider these parameters 
constants and, hence, we will let them being local functions of the 
space-time coordinates. The $x$ dependence also removes the known 
arbitrariness in the CPT-odd $\,k^\mu$-term\footnote{For constant 
$k^\mu$ in Minkowski space-time, the field reparametrization 
$$ 
\ph (x) \,\to\,  \ph (x) \,.\, e^{i \bar{k}^\mu x_\mu} 
\,, \qquad \quad 
\ph^* (x) \,\to\, e^{-i \bar{k}^\mu x_\mu} .\, \ph^* (x) 
\,,
$$ 
with $\bar{k}^\mu = (\eta^{\mu\nu} + K^{\mu\nu})^{-1} k_\nu $ leads to 
a new theory without the CPT-odd $k^\mu j_\mu$-term, but with a new 
mass definition $m^2 \,\to\, m^2 + \bar{k}_\mu \bar{k}^\mu$.} 
\cite{GravityCPTL,Altschul:2006jj}. No one assumptions about the 
violating parameters are made\footnote{Of course, for the Lagrangian 
\eq{action} be real and a scalar quantity the parameter $K^{\mu\nu}$ 
must possess, in general, a real symmetric part plus an antisymmetric 
imaginary one.}. The role of Lorentz and CPT symmetry breaking terms 
in the scalar sector of SME was widely studied in the flat space-time 
limit. The first bounds on the symmetry-breaking terms for the Higgs 
field were obtained in \cite{Anderson:2004qi} and quantum loops 
effects were considered in Refs.~\cite{Ferrero:2011yu,Altschul:2012ig,Carvalho:2013wsa,Vieira:2015jxa,Nascimento:2017ebc}. 
For further effects in violating scalar field theories see also the 
applications in the Yukawa potential \cite{Altschul:2006jj}, effective 
potential \cite{BaetaScarpelli:2017uxu}, Casimir effect \cite
{Casimir}, defect structures \cite{Bazeia:2005tb} and Bose-Einstein 
condensates \cite{Casana:2011bv}.

In what follows we consider the calculation of one-loop 
divergences in two separate cases. First we consider the minimal 
$\xi=0$ theory with a real scalar field. The reason is because in this 
case the current $i \left( \ph^* \, \pa_\mu \ph - \ph \, \pa_\mu \ph^* 
\right)$ drops out and the tensor $K^{\mu\nu}$ is necessary a 
symmetric tensor. Therefore, for this simpler case it is possible 
through a redefinition of the metric tensor to obtain 
a closed answer for the counterterms which is valid to all 
orders in the Lorentz violation parameter. After that, we are going to 
consider the general case described by the full theory \eq{action}. For 
this more complicated case, we are going to restrict our calculations to the first
order in the symmetry-breaking parameters. 

\subsection{One-loop divergences: minimal coupling with gravity}
\label{sec-div-mim}

As a first example, consider the massive real scalar field minimally 
coupled with gravity. In this case, the whole expression for the 
action reads:
\beq
\n{act-min}
S &=& 
\frac12\, \int d^4 x \sqrt{-g}\, 
\left\{ \left(g^{\mu\nu}+ K^{\mu\nu}\right) 
\pa_\mu \ph \, \pa_\nu \ph
- m^2 \ph^2 \right\}
\,.
\eeq
In order to evaluate the one-loop divergences, let us define a new 
metric,
\beq
\n{Gmn}
G^{\mu\nu}
&=&
g^{\mu\nu} + K^{\mu\nu}\,.
\eeq
After that, the action \eq{act-min} becomes
\beq
\n{act-G}
S &=&
\frac12\, \int d^4 x \sqrt{-G} \, f (x) \,
\left( G^{\mu\nu} D_\mu \ph \, D_\nu \ph
- m^2 \ph^2 \right)\,,
\eeq
where $\,G = \det \big( G_{\mu\nu} \big)$ is the determinant of the metric 
$G_{\mu\nu}$, defined as the inverse to $\,G^{\mu\nu}$ and
\beq
f (x) = \sqrt{\frac{\det \big( g^{\mu\nu} + K^{\mu\nu} \big) }
{\det \big( g^{\mu\nu} \big)}} 
\eeq
is a new background scalar field. Also, here $\,D_\mu\,$ is the 
covariant derivative constructed with the affine connection
\beq
\n{Up}
\Upsilon^\tau_{\al\be}
&=&
\frac12\,G^{\tau\la}\big(\pa_\al G_{\la\be}
+ \pa_\be G_{\al\la}-\pa_\la G_{\al\be}\big)
\,,
\eeq
defined in terms of the new metric. In the following related calculations, 
the indexes are lowered and raised with $G_{\mu\nu}$ and with 
its inverse. It is also very useful to introduce the corresponding 
curvature tensor
\beq
\n{K}
\big[ D_\mu \,,\, D_\nu \big] \, A^{\al}
&=&
{\cal K }^{\al}_{\,.\,\,\be \mu \nu} \, A^{\be}
\eeq
and its contractions
$\,{\cal K }_{\al\be} = G^{\mu\nu}\,{\cal K}_{\mu\al\nu\be}\,$ and 
$\,{\cal K} = G^{\mu\nu}\, {\cal K}_{\mu\nu}$. These new curvatures 
differ from the usual Riemann, Ricci tensors and scalar curvature 
$\,R\,$ by terms of first and higher orders in the Lorentz violating 
parameter $K^{\mu\nu}(x)$.

The procedure described above is a known calculation method which is
commonly used in Lorentz violating real scalar field theory 
\cite{Altschul:2006jj,Ferrero:2011yu}. However, since here the Lorentz 
violating parameter $K^{\mu\nu}$ is not constant, the answer will not 
be given only in terms of $K^{\mu\nu}$ determinants, as it was in the 
flat space-times cases, but also in terms of the new 
curvature tensor ${\cal K }^{\al}_{\,.\,\,\be \mu \nu}\,$. 
The same method was also recently applied in 
Ref.~\cite{Buchbinder:2017zaa} for the quantization of the 
Stueckelberg scalar sector of massive vector field theory with 
nonminimal coupling with gravity. 

Starting from Eq.~\eq{act-G}, the divergences derivation becomes 
pretty much standard. The divergent part of the one-loop effective action 
is given by the expression
\beq
\n{EA-G}
\Ga^{(1)}_{div} &=& 
\frac{i}{2} \, \Tr \ln \left( D^2
+ 2 \, \hat{l}^\mu \, D_\mu - m^2 \right)\Big|_{div}\,,
\eeq
where
\beq
\hat{l}_\mu
&=& \frac{1}{2}\, \pa_\mu (\ln f)
\qquad
\mbox{and}
\qquad
D^2 = G^{\mu\nu} D_\mu D_\nu\,.
\n{D}
\eeq 
The expression in Eq.~\eq{EA-G} can be evaluated in curved space-time by 
means of the standard Schwinger-DeWitt technique \cite{dewitt}. 
According to this method, the algorithm for the calculation of 
one-loop divergences, in dimensional regularization, is 
\beq
\Ga^{(1)}_{div} \,=\,
- \frac{\mu^{n-4}}{\epsilon}\, \int d^n x \sqrt{-G} \,
\Big[ \frac{1}{180} ({\cal K}_{\mu\nu\al\be}^2 
- {\cal K}_{\al\be}^2 + D^2 {\cal K})
+ \frac12 \, \hat{P}^2_{min} + \frac16 \, D^2 \hat{P}_{min}
\Big]\,,
\eeq
where $\epsilon = (4 \pi)^2 (n - 4)$ is the dimensional regularization 
parameter, $\mu$ is the mass dimensional parameter of renormalization
and
\beq
\n{Pmin}
\hat{P}_{min}
&=&
- \,m^2 + \frac{1}{6}\, {\cal K} - D_\mu  \hat{l}^\mu
- \hat{l}_\mu \, \hat{l}^\mu
\,.
\eeq
From Eq.~\eq{Pmin} we obtain
\beq
\frac12\, \hat{P}_{min}^2
&=&
\frac12\, m^4 - \frac16 \, m^2 {\cal K} + \frac{1}{72}\, {\cal K}^2
- m^2 F + \frac16\,{\cal K}  F
+ \frac12\, F^2
\,,
\eeq
where in the last expression we introduced the useful new notation
\beq
F
&=&
-\,\frac{1}{\sqrt{f}}\,\big(D^2 \sqrt{f} \big)
\,.
\label{F}
\eeq
Thus, the one-loop divergences can be written in the form
\beq
\n{div-s-G}
\Ga^{(1)}_{div} &=& 
- \,\frac{\mu^{n-4}}{\epsilon} \, \int d^n x \sqrt{-G} \
\left\{
 \frac{1}{180} \, {\cal K}_{\al\be\mu\nu}^2
- \frac{1}{180} \,{\cal K}_{\al\be}^2
+ \frac{1}{30} \, D^2 {\cal K}
+ \frac{1}{72} \, {\cal K}^2
\right.
\nonumber
\\
&& \left.
- \frac{1}{6} \, m^2 {\cal K}
+ \frac{m^4}{2}
+ \frac16 \, {\cal K} F
- m^2 F
+ \frac12 \, F^2
+ \frac16\, D^2 F
\right\}\,.
\eeq
The expression \eq{div-s-G} is the result of a standard QFT 
calculation in the theory with the new background metric 
$\,g^{\mu\nu}+K^{\mu\nu}\,$. In terms of this new metric formula 
\eq{div-s-G} has a rather standard form. At the same time, in terms of the 
original fields, $\,g_{\mu\nu}\,$ and $\,K_{\mu\nu}$, the divergences 
are given by an infinite series expression. This is an expected result 
which is corroborated by power counting based arguments, since the 
Lorentz violating parameter $K^{\mu\nu}$ is dimensionless.

The Eq.~\eq{div-s-G} enables us to obtain the one-loop divergences in 
terms of the original metric $g^{\mu\nu}$ in each desired order in the 
Lorentz violating parameter $K^{\mu\nu}(x)$. To obtain the explicit 
expression for the leading first order, we can use the standard 
expansions
\beq
G_{\mu\nu}
&=&
g_{\mu\nu} - K_{\mu\nu} + \dots
\,,
\qquad \quad
\sqrt{-G}
=
\sqrt{-g}\, \Big( 1 - \frac12\, K + \dots \Big)
\,,
\\
\nonumber
{\cal K}_{\mu\nu\al\be}
&=&
R_{\mu\nu\al\be}
+ \frac12 \, \big( \na_\mu \na_\al K_{\be\nu}
- \na_\nu \na_\al K_{\be\mu}
+ \na_\nu \na_\be K_{\al\mu}
- \na_\mu \na_\be K_{\al\nu}
\\
&+& R^\rho_{\,.\,\al\mu\nu} \, K_{\rho\be}
- R^\rho_{\,.\,\be\mu\nu} \, K_{\rho\al}
\big) + \dots
\,,
\eeq
where $\,K = g^{\mu\nu} K_{\mu\nu}$. For a more detailed exposition of 
the first-order formulae see, e.g., \cite{Buchbinder:2017zaa}.
Using those expansions, we can find the one-loop divergences written 
down in terms of original metric $\,g_{\mu\nu}\,$ in the first order 
in the Lorentz violating parameter
\beq
\n{fin-act-min}
\hspace{-6.2cm}
\Ga^{(1)}_{div} &=& 
-\frac{\mu^{n-4}}{\epsilon}\,
\int d^n x \sqrt{-g} \,
\Big\{\frac{1}{60}\, R_{\mu\nu} \, \Box K^{\mu\nu} 
- \frac{1}{30}\, R\, \na_\mu \na_\nu K^{\mu\nu}
+ \frac{1}{90} \,K^{\mu\nu}\, R^{\al\be}\, R_{\al\mu\be\nu}
\nonumber
\\
\hspace{-0.2cm}
&+& \frac{1}{90}\, K^{\mu\nu}\, R_{\mu\rho\al\be} \, 
R^{\,.\,\rho\al\be}_{\nu}
- \frac{1}{45} \,K^{\mu\nu}\, R_{\mu\al} \, R^{\al}_\nu
+ \frac{1}{36}\, K^{\mu\nu} \, R \, R_{\mu\nu}
- \frac{K}{2}\, \Big[
\frac{1}{180}\, R_{\mu\nu\al\be}^2 
\nonumber
\\
\hspace{-0.2cm}
&-& \frac{1}{180}\,R_{\mu\nu}^2 
+ \frac{1}{30}\, \Box R \,
+ \frac{1}{72}\, R^2 
+ \frac{m^2}{6}\, R  
+ \frac{m^4}{2} 
\Big]
+ \frac{m^2}{6}\,  K^{\mu\nu} \, R_{\mu\nu}
\Big\}
+ \Ga^{(1)}_{vac}[g_{\mu\nu}] 
\,,
\eeq
where
\beq
\n{sc-min-div}
\Ga^{(1)}_{vac}[g_{\mu\nu}]
&=&
- \,\frac{\mu^{n-4}}{\epsilon} \int d^n x \sqrt{-g} \,
\Big\{
\frac{1}{180}\, R_{\mu\nu\al\be}^2 
- \frac{1}{180}\,R_{\mu\nu}^2 
+ \frac{1}{30}\, \Box R
+ \frac{1}{72} \, R^2
\nonumber
\\
&+& \frac{m^2}{6} \,  R
+ \frac{m^4}{2} \,
\Big\}
\eeq
is the divergent part of the pure metric dependent vacuum effective 
action of minimally coupled scalar field (see, e.g., 
\cite{book,birdav}). For the sake of brevity, in Eq.~\eq{fin-act-min} 
(and in the following formulas) we disregarded the total derivative 
terms in the Lorentz- and/or CPT-violating sector. 

\subsection{One-loop divergences: nonminimal coupling with gravity}
\label{sec-div-nmim}

Let us consider now the case of nonminimally coupled with gravity 
Lorentz and CPT violating complex massive scalar field with general 
nonminimal coupling parameter $\xi$. The action \eq{action} can be 
cast in the bilinear form 
\beq
S &=& - \frac12 \int d^4 x \sqrt{-g}
\, \left(\begin{array}{cc}
\ph  & \ph^* \end{array} \right)
\hat{\mathbf H}
\left(\begin{array}{cc}
\ph^*
\\
\ph \end{array} \right)\,,
\eeq
where due to the presence of the extra adimensional Lorentz violating 
parameter $K^{\mu\nu}(x)$, the differential bilinear operator 
$\,\hat{\mathbf H}\,$ has a nonstandard general nonminimal structure, 
namely, 
\beq
\n{bilii}
\hat{\mathbf H} &=& \hat{\mathbf H}_m + \hat{\mathbf H}_{nm}
\,,
\eeq
where
\beq
\hat{\mathbf H}_m \,=\, \hat{\bf 1} \Box
+ 2 \, \hat{\bf L}^{\mu} \, \na_\mu
+ \hat{\bf \Pi}
\,
\eeq
is the minimal part of bilinear operator in quantum fields and
\beq
\hat{\mathbf H}_{nm} \,=\, \hat{\bf K}^{\mu\nu} \na_\mu \na_\nu
\eeq
is the nonminimal part. The relevant matrices are defined by
\beq
\hat{\bf 1}
&=&
\left(\begin{array}{cc}
1 & 0 \\
0 & 1
\end{array} \right) \,,
\nonumber
\\
\hat{\bf L}^\mu
&=&
\left(\begin{array}{cc}
\frac12\, \na_\nu K^{\mu\nu} + i k^\mu & 0 \\
0 & \frac12\, \na_\nu K^{\mu\nu} - i k^\mu
\end{array} \right) \,,
\nonumber
\\
\hat{\bf \Pi}
&=&
 \left(\begin{array}{cc}
m^2 - \xi R & 0 \\
0 & m^2 - \xi R 
\end{array} \right) \,,
\nonumber
\\
\hat{\bf K}^{\mu\nu}
&=&
\left(\begin{array}{cc}
K^{\mu\nu}  & 0 \\
0 & K^{\mu\nu}
\end{array} \right) \,.
\eeq
Here and in the following we use bold notations for the matrix 
operators only.

The one-loop divergent part of the effective action is then given by 
\beq
\n{EA}
\Ga^{(1)}_{div} \,=\, 
\frac{i}{2}\, \ln \Det \hat{ \mathbf H} \big|_{div} 
\,=\,
\frac{i}{2}\, \Tr \ln \hat{\mathbf H} \big|_{div}
\,.
\eeq
Our next purpose is derive the divergent expression \eq{EA} 
through heat kernel related calculations. However, the bilinear 
operator \eq{bilii} has a nonminimal form because of the presence of 
$K\na\na$-term, therefore, the standard Schwinger-DeWitt algorithm 
used before cannot be applied here. The formalism for dealing with 
nonminimal operators is the generalized Schwinger-DeWitt technique of 
Barvinsky and Vilkovisky \cite{bavi85}. Nevertheless, this well 
elaborated technique works only in the cases when nonminimality can be 
parametrized by some continuous parameter, in such away one can 
integrate over this parameter from zero, corresponding to the minimal 
limit, to any given value. In the case of Eq.\eq{bilii} one meets a 
tensor field and not just a parameter. Therefore, the known technique 
of dealing with nonminimal operators cannot be applied either.

In this situation, in order to work with such complicated operator we 
can follow the method developed in Ref.~\cite{CPTLTiSh} for analogous 
calculations in Lorentz/CPT violating electrodynamics. The main idea 
is to introduce the inverse of the minimal operator 
$\hat{\mathbf H}_m^{-1}$ and make the transformation
\beq
\n{exp}
\Tr \ln \hat{ \mathbf H} &=&
\Tr \ln ( \hat{\mathbf H}_m + \hat{\mathbf H}_{nm}) =
\Tr \ln \hat{\mathbf H}_m 
+ \Tr \ln ( 1 + \hat{\mathbf H}_m^{-1} \,.\, \hat{\mathbf H}_{nm})
\nonumber
\\
&=&
\Tr \ln \hat{\mathbf H}_m
+ \Tr \hat{\mathbf H}_{nm} \,.\, \hat{\mathbf H}_m^{-1}
- \frac12\, \hat{\mathbf H}_{nm} \,.\, \hat{\mathbf H}_m^{-1}
\,.\, \hat{\mathbf H}_{nm} \,.\, \hat{\mathbf H}_m^{-1}
+ \dots
\,,
\eeq
where we have used the basic properties of the 
logarithm and performed its power series expansion. Now, the first 
term of last line of Eq.~\eq{exp} contains only a minimal operator and 
can be directly calculated by the standard Schwinger-DeWitt technique 
\cite{dewitt}, while the rest of the expression \eq{exp} contains 
nonlocal nonminimal structures, which can be, in principle, reduced 
into the universal functional traces of Ref.~\cite{bavi85}.

Since the parameter $K^{\mu\nu}$ is dimensionless, it is possible to 
show, by power counting based arguments, that every term in the 
infinity series \eq{exp} gives contributions to the counterterms. The 
situation here is analogue to the quantum gravity on flat background, 
$\,g_{\mu\nu}\,=\,\eta_{\mu\nu}+h_{\mu\nu}\,$. Because the metric tensor
is also dimensionless, there is in this theory an infinite number of 
one-loop diagrams which are divergent. But, in the metric case there 
is the principle of general covariance, allowing to transform all such 
infinite divergent contributions into a small number of covariant 
invariant expressions in terms of the curvature tensors 
$R_{\mu\nu\al\be}\,$, $R_{\mu\nu}$ and $R$. In the case under 
consideration of Lorentz/CPT violating scalar field, there is no 
principle allowing to transform such infinite number of counterterms 
\eq{exp} into specific invariants constructed from $K^{\mu\nu}$. 
Unfortunately, there is no hope to find a closed solution, as in the 
previously minimal scalar case, and the series \eq{exp} must be 
truncated in some desired order. Since the Lorentz and CPT violating 
parameters are assumed to be very small, the calculations for general 
$\xi$ will be restricted to the first order in the symmetry-breaking 
terms. Then, the first order result of expression \eq{exp} is given by
\beq
\n{exp-1st}
\Tr \ln \hat{ \mathbf H} &=&
\Tr \ln \hat{\mathbf H}_m
+ 2 \Tr \hat{H}_{nm} \,.\, \hat{H}_0^{-1}
\,,
\eeq
where
\beq
\n{bili0}
\,\hat{H}_0 &=& 
\Box 
+ m^2 - \xi R
\,,
\eeq
is the standard bilinear operator for scalar field nonminimally 
coupled with gravity and
\beq
\n{Hnms}
\hat{H}_{nm} \,=\, K^{\mu\nu} \na_\mu \na_\nu 
\,.
\eeq

Let us now consider the evaluation of the divergences contained in 
expression \eq{exp-1st}. The first term in this formula possesses only 
minimal differential operators and it is possible to obtain the 
divergences, as before, by using the known formula of the 
Schwinger-DeWitt technique,
\beq
\n{div-min}
\frac{i}{2} \Tr \ln \hat{\mathbf H}_m \big|_{div}
\,=\,
- \frac{\mu^{n-4}}{\epsilon} \int d^n x \sqrt{-g}\, \tr 
\Big[ \frac{\hat{\bf 1}}{180} (R_{\mu\nu\al\be}^2 - R_{\al\be}^2)
+ \frac12 \hat{\bf P}^2 
\Big],
\eeq
where
\beq
\hat{ \bf P} = \hat{ \bf P}_0
- \na_\mu \hat{\bf L}^\mu + \dots
\,,
\qquad \mbox{with} \qquad
\hat{\bf P}_0 = \hat{\bf \Pi} + \frac16 \, \hat{\bf 1} \, R
\,.
\eeq

Then, up to the first order in the new parameters, we have
\beq
\n{nda}
\frac12 \,  \tr \hat{\bf P}^2
&=&
 \frac12 \, \tr \hat{ \bf P}_0^2
- \tr \hat{\bf P}_0 \, \na_\mu \hat{\bf L}^\mu 
+ \dots
\eeq
and also, 
\beq
\n{trP0L}
\tr \hat{\bf P}_0 \, \na_\mu \hat{\bf L}^\mu &=& 
\Big( \xi - \frac{1}{6} \Big)\, R \, (\na_\mu \na_\nu K^{\mu\nu}
+ i \na_\mu k^\mu - i \na_\mu k^\mu)
\\
&=& \Big( \xi - \frac{1}{6} \Big)\, R \, \na_\mu \na_\nu K^{\mu\nu}
\,.
\nonumber
\eeq
Formula \eq{trP0L} is the only source of contribution to divergences 
of the CPT-odd violating parameter $k^\mu(x)$. As explicitly shown this 
parameter gives no contribution to vacuum renormalization. A similar
situation occurs in the Lorentz and CPT-violating 
electrodynamics, where the CPT-odd $(k_{AF})^\mu$ parameter also do 
not contribute to pure vacuum counterterms 
\cite{CPTLGuiSh,CPTLSh,CPTLTiSh}. Just as in the QED case we expect 
that the odd parameters may contribute to the interaction theory. In 
order to understand this, we can remember that in the scalar 
electrodynamics, e.g., there is a mixing between the gauge field 
$A_\mu(x)$ with the scalar field $\ph(x)$ and the CPT-odd parameter
$k_\mu(x)$ through the gauge covariant derivatives 
$D_\mu \,=\, \na_\mu + i \,e\,A_\mu\,$, which are present in the 
current term $j_\mu$.

Finally, by the use of Eqs.~\eq{nda} and \eq{trP0L}, formula \eq{div-min} 
reduces to
\beq
\n{div-m}
\frac{i}{2} \, \Tr \ln \hat{ \mathbf H}_m \big|_{div}
\,=\,
- \frac{\mu^{n-4}}{\epsilon}\, \int d^n x \sqrt{-g} \,\,
\Big( \xi - \frac{1}{6} \Big)\, R \, \na_\mu \na_\nu K^{\mu\nu}
\,+\, \Ga^{(1)}_{vac}[g_{\mu\nu}] 
\,+\, \dots
\,,
\eeq
where $\Ga^{(1)}_{vac}[g_{\mu\nu}]$ is the divergent part of the 
metric dependent vacuum effective action of nonminimally complex 
scalar field
\beq
\n{sc-div-vac}
\Ga^{(1)}_{vac}[g_{\mu\nu}]
&=&
- \,\frac{2\mu^{n-4}}{\epsilon} \int d^n x \sqrt{-g} \,
\Big\{
\frac{1}{180}\, R_{\mu\nu\al\be}^2 - \frac{1}{180}\,R_{\mu\nu}^2 
- \frac16\, \Big(\xi - \frac{1}{5} \Big) \, \Box R \,
\nonumber
\\
&+& \frac{1}{2}\, \Big( \xi -\frac{1}{6} \Big)^2 R^2
- m^2\, \Big( \xi - \frac{1}{6} \Big)\, R
+ \frac{m^4}{2}
\Big\}
\,.
\eeq

For calculating the divergent part of the nonminimal piece of 
Eq.~\eq{exp-1st}, we first need invert the operator $\,\hat{H}_0\,$ 
and find its nonlocal expression. Up to the background dimension of 
$1/l^4$ (for introduction in this terminology, see Ref.~\cite{bavi85}) 
the inverse operator can be expressed as
\beq
\hat{H}_0^{-1} &=& \frac{1}{\Box}
+ \left( \xi R - m^2 \right) \, \frac{1}{\Box^2}
+ \left( m^4 -2 \xi m^2 R + \xi^2 R^2 - \xi \Box R \right) 
\frac{1}{\Box^3}
\\
&-& \,2 \, \xi \, (\na^\mu R) \, \na_\mu \frac{1}{\Box^3}
+ \,4 \, \xi \, (\na^\mu \na^\nu R) \, \na_\mu \na_\nu \, 
\frac{1}{\Box^4}
+ {\cal O} (l^{-5})
\,.
\nonumber
\eeq
The higher background dimension, $\,{\cal O}(l^{-5})\,$ terms, can be 
safely omitted here because they do not contribute to divergences 
\cite{bavi85}. Using equation \eq{Hnms} one can obtain the relation
\beq
\n{nmH0}
&&
\hspace{-1.2cm}
\Tr \hat{H}_{nm} \,.\, \hat{H}_{0}^{-1} \,=\, 
K^{\mu\nu} \, \na_\mu \na_\nu \,
\frac{1}{\Box}
+ \left( \xi R - m^2\right)
K^{\mu\nu} \, \na_\mu \na_\nu \,
\frac{1}{\Box^2}
+ \xi \, K^{\mu\nu} \, (\na_\mu \na_\nu R)
\, \frac{1}{\Box^2}
\nonumber
\\
&&
\hspace{-1.2cm}
\,-\, 4 \, \xi \, K^{\mu\al}\, (\na_\al \na^\nu R) 
\, \na_\mu \na_\nu \frac{1}{\Box^3}
+ \left( m^4 - 2m^2\, \xi\, R + \xi^2 \, R^2 
- \xi \, \Box R \right) K^{\mu\nu} \,
\na_\mu \na_\nu \, \frac{1}{\Box^3}
\nonumber
\\
&&
\hspace{-1.2cm}
\,+\, 4 \, \xi \, (\na^\mu \na^\nu R) \, K^{\al\be}
\, \na_\al \na_\be \na_\mu \na_\nu \, 
\frac{1}{\Box^4} 
+ {\cal O} (l^{-5})
\,.
\eeq
Once more, in the last formula we do not write explicitly the 
${\cal O}(l^{-5})$ structures and also the functional traces with 
dimensionality $l^{-3}$, because they are irrelevant to the 
divergences.

Expression \eq{nmH0} is already in the form that allows us to apply 
the tables of universal functional traces of Ref.~\cite{bavi85}. 
The calculation is straightforward and the intermediary results are 
shown in Appendix \ref{apA}. The final result has the form
\beq
\n{div-nm}
\hspace{-0.7cm}
\frac{i}{2}\, \Tr \hat{H}_{nm} \,.\, \hat{H}_{0}^{-1} \big|_{div}
&=& 
-\frac{\mu^{n-4}}{\epsilon}\,
\int d^n x \sqrt{-g} \,
\Big\{
\frac{1}{60}\, R_{\mu\nu} \, \Box K^{\mu\nu} 
+ \Big( \frac{1}{20} - \frac13 \, \xi \Big)\, 
R\, \na_\mu \na_\nu K^{\mu\nu} 
\nonumber
\\
&+& \frac{1}{90} \,K^{\mu\nu}\, R^{\al\be}\, R_{\al\mu\be\nu}
+ \frac{1}{90}\, K^{\mu\nu}\, R_{\mu\rho\al\be} \, 
R^{\,.\,\rho\al\be}_{\nu}
- \frac{1}{45} \,K^{\mu\nu}\, R_{\mu\al} \, R^{\al}_\nu
\nonumber
\\
&-& \frac{1}{6}\, \Big( \xi - \frac{1}{6} \Big) 
\, K^{\mu\nu} \, R \, R_{\mu\nu}
-\frac{K}{2} \,\Big[\,\frac{1}{180}\, R_{\mu\nu\al\be}^2 
- \frac{1}{180}\,R_{\mu\nu}^2 
\\
&-& \frac16\, \Big(\xi - \frac{1}{5} \Big) \, \Box R \,
+ \frac{1}{2}\, \Big( \xi -\frac{1}{6} \Big)^2 R^2 
-\, m^2\, \Big(\xi -\frac16 \Big)\, R  
\nonumber
\\
&+& \frac{m^4}{2}
\,\Big]
+ \frac{m^2}{6}\, K^{\mu\nu} R_{\mu\nu}
\Big\}
\,.
\nonumber
\eeq

Finally, from equations \eq{EA}, \eq{exp-1st}, \eq{div-m} and 
\eq{div-nm} we arrive at the result for the one-loop divergences 
in the first order in Lorentz violating parameter 
\beq
\n{div-comp}
\Ga^{(1)}_{div} &=& 
-\frac{2\mu^{n-4}}{\epsilon}\,
\int d^n x \sqrt{-g} \,
\Big\{
\frac{1}{60}\, R_{\mu\nu} \, \Box K^{\mu\nu} 
+ \frac16\, \Big( \xi - \frac15 \Big)
\, R\, \na_\mu \na_\nu K^{\mu\nu} 
\nonumber
\\
&+& \frac{1}{90} \,K^{\mu\nu}\, R^{\al\be}\, R_{\al\mu\be\nu}
+ \frac{1}{90}\, K^{\mu\nu}\, R_{\mu\rho\al\be} 
\, R^{\,.\,\rho\al\be}_{\nu}
- \frac{1}{45} \,K^{\mu\nu}\, R_{\mu\al} \, R^{\al}_\nu
\\
&-& \frac{1}{6}\, \Big( \xi - \frac{1}{6} \Big) \, K^{\mu\nu} \, R 
\, R_{\mu\nu}
- \frac{K}{2}\,\Big[\,\frac{1}{180}\, R_{\mu\nu\al\be}^2 
- \frac{1}{180}\,R_{\mu\nu}^2 
- \frac16\, \Big(\xi - \frac{1}{5} \Big) \, \Box R \,
\nonumber
\\
&+& \frac{1}{2}\, \Big( \xi -\frac{1}{6} \Big)^2 R^2 
- m^2\Big(\xi -\frac16 \Big) \, R  
+ \frac{m^4}{2}
\,\Big]
+ \frac16\, m^2 \, K^{\mu\nu} \, R_{\mu\nu}
\Big\}
+ \Ga^{(1)}_{vac}[g_{\mu\nu}]
\nonumber 
\,.
\eeq

Let us start the analysis of our result. First of all, we can verify 
that in the $\,\xi\,=\,0\,$ limit we arrive exactly at the same result 
for the first order one-loop divergences in the minimal case 
Eq.~\eq{fin-act-min} obtained by the previously 
calculation method\footnote{The extra $2$ factor is because the theory 
of $N$ complex scalar fields can also be written in terms of a model 
with $2N$ real scalar fields.}. Second, the vacuum part 
\eq{sc-div-vac} has the well known standard form of the divergences in 
curved space-time for the massive scalar field theory nonminimally 
interacting with gravity. This is perfectly consistent with general 
features of renormalization in curved space-time, because the 
semiclassical renormalizable theory always includes higher derivative 
terms in the gravitational sector (see, e.g., \cite{book,birdav}). In 
our case, this also means that the renormalization of the nonviolating 
sector is performed independently
on the external Lorentz/CPT 
symmetry-breaking fields. In the case when some of the violating 
fields are present, the consistent form of the vacuum action becomes 
much more complicated and involves the dependence on these extra 
fields. Our one-loop calculations show which terms can emerge as 
counterterms in the scalar field case. Therefore, the minimal set of 
structures which are requested by renormalizability in the 
gravitational action can be expressed as
\beq
\n{arel}
S_{grav} &=& \int d^4 x \sqrt{-g} \,
\left\{ v(x) + u(x)\, R  + s^{\mu\nu}(x)\, R_{\mu\nu}
\right\} + S_{HD}
\,,
\eeq
where the last part $\,S_{HD}\,$ represents the generalized higher 
derivative term
\beq
\n{LCHD}
\hspace{-0.7cm}
S_{HD}
&=& \int d^4 x \sqrt{-g} \,
\big\{\Phi_1(x) \,R_{\mu\nu\al\be}^2
+ \Phi_2(x) \,\,R_{\mu\nu}^2 
+ \Phi_3(x) \, \Box R
+ \Phi_4(x) \, R^2
\\
\hspace{-0.7cm}
&+& \zeta_1^{\mu\nu}(x) \, R_{\mu\rho\al\be} \, R_\nu^{\,.\,\rho\al\be}
+ \zeta_2^{\mu\nu}(x) \, R^{\al\be} \, R_{\mu\al\nu\be} 
+ \zeta_3^{\mu\nu}(x) \, R_{\mu\rho} \, R^\rho_{\nu} 
+ \zeta_4^{\mu\nu}(x) \, R \, R_{\mu\nu}
\nonumber
\big\} 
\,.
\eeq 
Let us notice that the terms 
$\,t^{\mu\nu\al\be}(x) \, R_{\mu\nu\al\be}\,$, 
$\,\eta_1^{\mu\nu\al\be}(x) \, R_{\rho\om\mu\nu} \, 
R^{\rho\om}_{\,.\,.\,\al\be}\,$, 
$\,\eta_2^{\mu\nu\al\be}(x) \, R_{\mu\rho}
\, R^\rho_{\,.\,\nu\al\be} \,$, 
$\,\eta_3^{\mu\nu\al\be}(x) \, R \, R_{\mu\nu\al\be}\, $ 
and 
$\,\eta_4^{\mu\nu\al\be}(x) \, R_{\mu\al} \, R_{\nu\be}\,$ 
which are necessary for the renormalization of Lorentz violating 
electrodynamics do not appear here in the scalar field case. Indeed this is
expect due to the different tensorial properties of the
symmetry-breaking fields in these two theories. Of course, this is 
also for the reason that we work only in first order in the 
Lorentz/CPT violating parameters and we expect that such structures 
will emerge in higher order calculations. In principle, the nonlinear 
terms can also be derived from the general expansion \eq{exp}, 
however, the derivation of such complicated functional traces will 
require significant efforts. 

The most remarkable aspect of the result \eq{div-comp} is that the 
cosmological constant-like divergence $m^4$ appears multiplied by 
a coefficient $K$ which may be coordinate-dependent. This means that 
in the theory where Lorentz-violating parameter $K^{\mu\nu}$ is not 
a constant, the cosmological constant cannot be constant, but should 
have some coordinate dependence. It would be certainly interesting to 
derive the upper bound for the time-dependence of $K(x)$ from 
laboratory experiments and compare it to the bounds for variable 
vacuum energy density in cosmology.
Also, many of the other new structures present in Eqs.~\eq{arel} and \eq{LCHD} can 
imply in new gravitational physical effects.
The investigation of the 
possible phenomenological manifestations of terms linear in curvature
was performed in Ref.~\cite{Bailey:2006fd} on the basis of PPN 
formalism and, recently, an extensive systematic analysis of the 
Lorentz violating higher derivatives terms has been started 
in \cite{HGLCPTNew}, also in the weak gravitational field 
approximation. According to \cite{HGLCPTNew} the presence of higher 
derivative violating terms leads to a modified Poisson equation for the 
gravitational potential in the form
\beq
\n{new-modified}
\De \ph_{grav.} (r) \,=\, - 4 \pi G \rho(r) 
+ (k_{eff}^{ijlk}) \, \pa_i \pa_j \pa_l \pa_k \ph_{grav.}(r) 
\,,
\eeq
which implies in diverse new phenomenological consequences. In formula 
\eq{new-modified} the violating parameter $\,k_{eff}^{ijlk}\,$ is 
constructed on the basis of the symmetry-breaking fields present in 
the higher derivative sector of gravitational action. Additionally, in 
Ref.~\cite{Hernaski:2014lga} the role of some Lorentz violating higher 
derivative terms was analyzed in the quantum gravity framework. Since 
action \eq{arel} is requested by the renormalization of Lorentz/CPT 
violating SME matter sector, the detailed analysis of all contained 
structures deserve a especial attention in both, classical and quantum 
levels.

\section{Local conformal symmetry and conformal anomaly}
\label{sec-CS}

It is pretty well known that the classical action of free scalar field 
theory in curved space-time is invariant, in the $m \,=\,0$ and
$\xi \,=\, 1/6\,$ limit, under the following transformations:  
\begin{eqnarray}
\n{tc1}
g_{\mu\nu} \,\to\, g'_{\mu\nu}\,=\, g_{\mu\nu}\,.\,e^{2\si(x)}
\qquad \mbox{and} \qquad
\ph \,\to\, \ph' \,=\, \ph \,.\,e^{-\si(x)}
\,.
\end{eqnarray}
The formula \eq{tc1} is called local conformal transformation and the 
corresponding action invariance is known as local conformal symmetry. 
The form of the Noether identity corresponding to this symmetry, in 
the on-shell limit\footnote{For Eq.~\eq{Ncf} to be valid in the 
Lorentz/CPT violating theories the symmetry-breaking parameters must 
obey their own dynamical equations. As discussed in \cite{GravityCPTL} 
this can be achieved if the violating fields originates from some 
spontaneous symmetry breaking mechanism.}, is  
\beq
\n{Ncf}
2 g_{\mu\nu} \, \frac{\de S}{\de g_{\mu\nu}} \,=\, 0
\,,
\eeq
which is interpreted as the vanishing trace of energy-momentum tensor 
$\,T^\mu_\mu \,=\, 0\,$. It is very important to note that the 
classical action of scalar field with Lorentz and CPT symmetry 
breaking terms \eq{action} also possesses local conformal invariance 
in the aforementioned limit, if we allow the Lorentz and/or 
CPT-violating parameters transform according to 
\beq
K^{\mu\nu} \,\to\,
K'^{\mu\nu} \,=\, K^{\mu\nu}\,.\,e^{-2\si(x)}
\,,
\quad \quad
k^\mu \,\to\, k'^\mu \,=\, k^\mu \,.\,e^{-2\si(x)}
\,.
\eeq

The breaking of equation \eq{Ncf} occurs only at quantum level because 
of the renormalization procedure. Such phenomenon is known as 
conformal anomaly or, simply, trace anomaly \cite{Duff94}. At quantum 
level the classical action of vacuum has to be replaced by the 
renormalized effective action
\beq
\Ga_R \,=\, S + \Ga^{(1)} + \De S
\,,
\eeq
where $\,\Ga^{(1)} \,=\, \Ga^{(1)}_{div} + \Ga^{(1)}_{fin}\,$ is the 
naive one-loop quantum correction to the classical action $S$ and 
$\De S\,$ is a local counterterm which is requested to cancel the 
divergent part of $\Ga^{(1)}$. The counterterm $\De S$ is the only 
source of nonconformal invariance of the effective action, because 
both the classical action and direct quantum contribution are 
conformal invariant. Thus, the expectation value of the trace 
$\langle T_\mu^\mu \rangle$ differs from zero and can be expressed by
\beq
\n{con-an}
\langle T_\mu^\mu \rangle \,=\,
- \frac{2}{\sqrt{-g}} \,g_{\mu\nu}\,
\frac{\de \Ga_R}{\de g_{\mu\nu}}
\Bigg|_{n=4} \,=\,
- \frac{2}{\sqrt{-g}}\, g_{\mu\nu}\,
\frac{\de \De S}{\de g_{\mu\nu}}
\Bigg|_{n=4}
\,.
\eeq
The form of the counterterm $\De S \,=\, -  \Ga^{(1)}_{div}\,$ for the 
conformal version of theory \eq{action} can be obtained from 
Eq.~\eq{div-comp}. The answer is  
\beq
\n{count}
\De S \,=\, 
\frac{2\mu^{n-4}}{\epsilon}\,
\int d^n x \sqrt{-g} \, L \left(g_{\mu\nu},K_{\mu\nu} \right)
\,-\, \Ga^{(1)}_{vac}[g_{\mu\nu}] 
\,,
\eeq
where
\beq
\n{L}
L \left(g_{\mu\nu},K_{\mu\nu} \right) &=&
\frac{1}{60}\, R_{\mu\nu} \, \Box K^{\mu\nu} 
- \frac{1}{180}\, R\, \na_\mu \na_\nu K^{\mu\nu} 
+ \frac{1}{90} \,K^{\mu\nu}\, R^{\al\be}\, R_{\al\mu\be\nu}
\nonumber
\\
&+& \frac{1}{90}\, K^{\mu\nu}\, R_{\mu\rho\al\be} \, 
R^{\,.\,\rho\al\be}_{\nu}
- \frac{1}{45} \,K^{\mu\nu}\, R_{\mu\al} \, R^{\al}_\nu
- \frac{K}{2}\, \Big( \,\frac{1}{180}\, R_{\mu\nu\al\be}^2 
\\
&-& \frac{1}{180}\,R_{\mu\nu}^2 
+ \frac{1}{180}\, \Box R \,
\Big)
\nonumber
\eeq
and
\beq
\n{sc-div-vac-con}
\Ga^{(1)}_{vac}[g_{\mu\nu}]
&=&
- \,\frac{2\mu^{n-4}}{\epsilon} \int d^n x \sqrt{-g} \,
\left\{
\frac{1}{120}\, C^2 - \frac{1}{360}\,E + \frac{1}{180} \, \Box R
\right\}
\,.
\eeq
In the above formula 
$C^2 = C_{\mu\nu\al\be} C^{\mu\nu\al\be} = R_{\mu\nu\al\be}^2 - 2 
R_{\mu\nu}^2 + \frac{1}{3}\, R^2$ 
is the square of Weyl tensor
and $E = R_{\mu\nu\al\be}^2 - 4 R_{\mu\nu}^2 + R^2\,$ is the integrand 
of the Gauss-Bonnet topological term (Euler density in $n=4$).

The calculation of expression \eq{con-an} with the counterterm 
\eq{count} can be done by many different ways 
\cite{birdav,Christensen,Deser:1976yx,Duff:1977ay,Asorey:2003uf}. 
Following Ref.~\cite{ConfPo}, the simplest one is by using the 
conformal parameterization of the metric,
\beq
\n{metric-t}
g_{\mu\nu} \,=\, g'_{\mu\nu} \,.\,e^{2\si(x)} \,,
\eeq
and by the direct application of the chain rule 
\beq
\n{rel}
- \frac{2}{\sqrt{-g}} \,g_{\mu\nu}
\, \frac{\de A[g_{\mu\nu}]}{\de g_{\mu\nu}}
\,=\, - \frac{1}{\sqrt{-g'}} \,e^{-4n\si}\,
\frac{\de A[g'_{\mu\nu} \,e^{2\si}]}{\de \si}
\Bigg|_{g'_{\mu\nu}\to g_{\mu\nu}\,,\,
\,\si \to 0}
\,,
\eeq
which is valid for any functional $A = A[g_{\mu\nu}]$. This procedure 
can be seen as a purely technical one and the form of the metric 
\eq{metric-t} is discarded after the anomaly derivation. 

One of the key parts of this general procedure are the conformal 
transformation rules of each quantity present in Eq.~\eq{count}. 
Besides the pure curvature terms, whose transformation rules can be 
found elsewhere \cite{Carneiro:2004rt}, we also need the 
transformation rule for the new Lorentz violating term \eq{L}. 
In fact, in the four dimensional space-time this term is conformal 
invariant
\beq
\n{conI}
\int d^4 x \sqrt{-g'} \,L \left(g'_{\mu\nu},K_{\mu\nu}' \right)
&=& \int d^4 x \sqrt{-g}\, L \left(g_{\mu\nu},K_{\mu\nu} \right)
\,.
\eeq
For the convenience of the reader we present the proof of 
Eq.~\eq{conI} in Appendix \ref{apB}. Indeed, the conformal symmetry 
\eq{conI} of the quantum correction in four dimensional space-time 
limit is expect to hold for conformal theories based on general 
standard arguments, thus, the direct algebraic proof of Eq.~\eq{conI} 
can also be seen as a test of verification of the cumbersome 
calculations which led to the answer \eq{div-comp}. On the top of 
that, formula \eq{conI} also implies that in the generic space-time 
with $n$ dimensions, the generalized $n$-dimensional form of 
Eq.~\eq{conI} gains a global $e^{(n-4)\si}$ multiplicative factor, 
besides some possible extra terms with derivatives of $\si(x)$.  All 
other expressions of our interest have the same general structure with 
the multiplicative exponential factor, and the non exponential terms 
are irrelevant due to the limit procedure in Eq.~\eq{rel}. 
Consequently, the application of identity \eq{rel} becomes simple.
Thus, one can find the final answer for the conformal anomaly,
\beq
\n{ano}
\langle T_\mu^\mu \rangle
&=& 
- \big[\, w C^2 + b E + c \Box R + 2 L(g_{\mu\nu},K_{\mu\nu})\, \big]
\,,
\eeq
where the parameters  \ $w$, $b$, $c$ \ are, in the complex scalar 
field case,
\beq
\left( w,b,c \right) &=& 
\frac{2}{(4\pi)^2}\,
\Big(\frac{1}{120}
\,,
-\frac{1}{360}
\,,
\frac{1}{180}
\Big)
\,.
\eeq
In the case of local conformal invariance there is always a well-known 
ambiguity in the value of $c$-parameter 
\cite{birdav,Duff94,Asorey:2003uf}. In a simplified way, the 
qualitatively net result is that this ambiguity is always equivalent 
to the freedom to add the local $R^2$-term to the classical action, 
since
\beq
\n{trDR2}
- \frac{2}{\sqrt{-g}}\, g_{\mu\nu} 
\frac{\de}{\de g_{\mu\nu}}
\int d^4 x \sqrt{-g} \, R^2
\,=\, 12 \Box R
\,.
\eeq
For more details in this subject the reader is referred to 
\cite{Asorey:2003uf}, where this issue was addressed with all 
technicalities.

\section{Anomaly-induced effective action}
\label{sec-aIEA}

One can use the conformal anomaly \eq{ano} to construct a differential 
equation for the finite part of the one-loop correction to the 
effective action
\beq
\n{dif-e}
\frac{2}{\sqrt{-g}} \, g_{\mu\nu}
\, \frac{\de \Ga_{ind}}{\de g_{\mu\nu}}
&=& w C^2 + b E + c \Box R + 2 L
\,.
\eeq

The solution of Eq.~\eq{dif-e} is known as anomaly-induced effective 
action. The integration of conformal anomaly is by the technical side 
not very difficult in the usual theory without the Lorentz violating 
term \cite{Ano-int} and it remains identically simple when this term 
is present \cite{CPTLTiSh}. The reason is because the new violating 
term \eq{L} possesses the same conformal properties of the square of Weyl 
tensor, which makes its inclusion a very simple exercise.

The anomaly-induced effective action can be presented in the simplest 
way by a noncovariant form, or in a more complicated one which is 
covariant and nonlocal. Additionally, by the introduction of auxiliary 
fields it can also be cast into a dynamically equivalent local and 
covariant form. Let us start from the simplest case and parameterize 
the metric tensor as in \eq{metric-t}, separating its conformal factor 
$\,\si(x)\,$. After that, we can rewrite Eq.~\eq{dif-e} using the 
relation \eq{rel} and the conformal transformation rules 
\cite{Carneiro:2004rt}
\begin{equation}
\sqrt{-g} \, C^2 \,=\, \sqrt{-g'} \, C'^2  
\,,
\end{equation}
\begin{equation}
\n{GBt}
\sqrt{-g} \, (E + \tfrac{2}{3} \Box R)  \,=\ 
\sqrt{-g'} \, (E' + \tfrac{2}{3} \Box' R' + 4 \De_4' \si)
\,,
\end{equation}
\begin{equation}
\sqrt{-g} \, \De_4  \,=\ \sqrt{-g'} \, \De'_4
\end{equation}
together with the Lorentz violating term transformation, 
Eq.~\eq{conI}. Here and below the quantities with primes are 
constructed using only the metric $g'_{\mu\nu}$. In particular, 
in the above formula $\De_4$ is the Paneitz operator \cite{Paneitz}
\beq
\De_4 \,=\, \Box^2 + 2 R^{\mu\nu} \na_\mu \na_\nu
- \frac{2}{3} R \, \Box + \frac{1}{3}\, (\na^\mu R) \na_\mu\,,
\eeq
which is a covariant, fourth derivative, self-adjoint and conformal 
invariant operator when acting on dimensionless scalar fields.

After the described procedure is completed, the formula \eq{dif-e} 
becomes very simple and integration in the $\si$ variable is 
straightforward. The solution for the effective action is
\beq
\n{ano-action}
\Ga_{ind} &=& S_c[g_{\mu\nu}',K_{\mu\nu}'] 
\,+\, \int d^4 x \sqrt{-g'} \,
\Big\{ w \si C'^2 + b \si ( E' - \tfrac{2}{3}\, \Box' R') 
+ 2 b \si \De'_4 \si
\nonumber
\\
&+&
2 \si L(g'_{\mu\nu},K'_{\mu\nu})
-\frac{3c+2b}{36} \, [R'-6(\na' \si)^2
- 6 \Box' \si]^2  \Big\}
\,,
\eeq
where $S_c = S_c[g_{\mu\nu},K_{\mu\nu}]$ is an unknown conformal 
invariant functional, which serves as an integration constant for 
Eq.~\eq{dif-e} and cannot be uniquely defined in the present scheme. 
Since it is a conformal invariant functional it does not depend on the 
conformal factor of the metric and is irrelevant for the dynamics of 
the metric in simple cases, as the cosmological 
Friedmann-Lema\^{i}tre-Robertson-Walker (FLRW) metric. However, when 
the violating fields are present, the automatic irrelevance of this 
term in the zero-order cosmology does not hold, because it depends on 
the Lorentz violating parameter and contributes for its dynamical 
equation. 

At the same time, this conformal invariant term can be ignored as a 
good approximation. The reason is because $\,S_c\,$ contains only 
sub-leading quantum corrections, while the rest of action 
\eq{ano-action} contains all the leading logarithm corrections 
\cite{Shapiro:2008sf}, with full information about the ultraviolet 
limit of theory. Moreover, the results obtained without this conformal 
term provide a very nice match with the answers obtained by other 
methods, as in the gravitational waves \cite{star83,waveana,HHR} and 
black holes \cite{BHano} cases. 

The solution \eq{ano-action} is noncovariant, in the sense that it is not 
written in terms of the original metric $g_{\mu\nu}$. In order to 
obtain the covariant solution we can use the following formula which 
is true for any conformal functional 
$A[g_{\mu\nu}] \,=\, A[g'_{\mu\nu}]$ 
\beq
\n{IGB}
&& \hspace{-1.5cm}
2 g_{\mu\nu} (y) \, \frac{\de}{\de g_{\mu\nu} (y)} \,
\int d^4 x \sqrt{-g(x)} \, A \left(E - \tfrac{2}{3} \,\Box R \right)
\nonumber
\\
&& 
\,=\, \frac{\de}{\de \si (y)} \,
\int d^4 x \sqrt{-g'(x)} \, A' \left(E' - \tfrac{2}{3} \,\Box' R'
+ 4 \De_4' \si \right) 
\Big|_{g'_{\mu\nu}\to g_{\mu\nu}\,,\,
\,\si \to 0}
\\
&& 
\,=\, 4 \, \sqrt{-g'(y)} \, \De'_4 A'
\,=\, 4 \, \sqrt{-g(y)} \, \De_4 A
\,. \nonumber
\eeq
In the above equation we have used the identity \eq{rel} together with 
the transformation rule \eq{GBt}. Introducing the Green function for 
the Paneitz operator,
\beq
\sqrt{-g(x)} \, \De_{4,x} G(x,y) \,=\, \de^4(x-y)
\eeq
and by means of relation \eq{IGB}, solving Eq.~\eq{dif-e} becomes 
direct. For example, for the Weyl squared term 
\beq
&&
\hspace{-1cm}
\, 2g_{\mu\nu}\, \frac{\de}{\de g_{\mu\nu}(y)} 
\int d^4 x \sqrt{-g(x)} \int d^4 y \sqrt{-g(y)} \,\,
C^2(x) \, \, G(x,y) \, \left( E - \tfrac{2}{3} \, \Box R \right)_y
\\
\nonumber
&&
\,=\, 4 \int d^4 x \sqrt{-g(y)} \, \De_{4,y}  G(x,y) \, C^2(x)
\,=\, 4 \, C^2(y)
\,.
\eeq
Using an analogous consideration for the other terms\footnote{For the 
$\Box R$-term in trace anomaly we can straightforward use formula 
\eq{trDR2} and find the local piece of anomaly-induced EA, 
Eq.~\eq{EAc}.} in \eq{dif-e}, we arrive at the solution
\beq
\n{EAnl}
\Ga_{ind} \,=\, S_c + \Ga_w + \Ga_b + \Ga_c
\,,
\eeq
where 
\beq
\n{EAw}
\Ga_w &=& \frac{1}{4} \, 
\int d^4 x \sqrt{-g(x)} \int d^4 y \sqrt{-g(y)} \,\,
\left(w C^2 + 2L\right)_x  G(x,y)  \left( E - \tfrac{2}{3} 
\, \Box R \right)_y
\,,
\\
\Ga_b &=& \frac{b}{8} \, 
\int d^4 x \sqrt{-g(x)} \int d^4 y \sqrt{-g(y)} \,\,
\left( E - \tfrac{2}{3} \, \Box R \right)_x  G(x,y) 
 \left( E - \tfrac{2}{3} \, \Box R \right)_y
\eeq
and
\beq
\n{EAc}
\Ga_c \,=\,\,-\, \frac{3c+2b}{36}\,\int d^4 x \sqrt{-g(x)}\,R^2(x)
\,.
\eeq
One can note that the Lorentz violating terms show up only in the 
first nonlocal term, Eq.~\eq{EAw}. 

At the next stage, the nonlocal expressions for the anomaly-induced EA 
can be presented in a local form through the introduction of two 
auxiliary scalar fields $\phi(x)$ and $\psi(x)$ \cite{Shapiro:1994ww} 
(the simpler one scalar form was known from much earlier, see 
\cite{Ano-int}). This procedure was discussed in details in 
Ref.~\cite{Shapiro:1994ww} and revised in 
\cite{ConfPo,Shapiro:2008sf}, so let us give just a final result for 
the local form of the anomaly-induced effective action,
\beq
\n{finaction}
\hspace{-1cm}
\Ga_{ind} &=& S_c[g_{\mu\nu},\,K_{\mu\nu}]
\,-\, \frac{3c+2b}{36}\,\int d^4 x \sqrt{-g(x)}\,R^2(x)
\,+\,  \int d^4 x \sqrt{-g(x)}\,\Big\{
\tfrac12 \,\phi \De_4\phi
\nonumber
\\
&-& \tfrac12 \,\psi\De_ 4\psi
+ \phi\left[
k_1\,
\left( C^2 + \tfrac{2}{w}\, L \right)
+\,k_2\,\left(E -\tfrac23\,{\Box}R\right)\,
\,\right]
+
l_1\,\psi\left(C^2
+  \tfrac{2}{w} \, L\right) \,\Big\}
\,,
\eeq
where
\beq
k_1 = - l_1 = - \frac{w}{2\sqrt{-b}}
\qquad \mbox{and} \qquad
k_2 = \frac{\sqrt{-b}}{2} 
\,.
\eeq
At the classical level the local covariant form \eq{finaction} is 
dynamically equivalent to the nonlocal covariant one \eq{EAnl}, which 
means that after solving the field equations for the fields $\phi(x)$ 
and $\psi(x)$ and plugging back these solutions in action 
\eq{finaction}, we come back to the previous formula \eq{EAnl}. 
The use of the local covariant form with auxiliary scalars is 
advantageous because the initial value problem for these fields are 
equivalent to the boundary conditions for the two Green functions 
present in the nonlocal covariant form \eq{EAnl}. By this reason, 
Eq.~\eq{finaction} is the most useful one for dealing with Hawking 
radiation from black holes \cite{BHano,Christensen:1977jc} or 
exploring the dynamics of gravitational waves on cosmological 
background \cite{waveana}. Also, the relevance to have two auxiliary 
fields instead of a single one field has been addressed in details 
in Refs.~\cite{BHano,Shapiro:1994ww}.

The actions \eq{ano-action}, \eq{EAnl}, \eq{finaction} represent the 
final product of our conformal anomaly integration. They correspond 
to the quantum correction to the classical gravitational action. In 
comparison with the previous standard case known
in literature, 
those formulas have extra Lorentz violating terms coming from the 
scalar field contribution. All information about the symmetry-breaking 
parameters is included in the $ L(g_{\mu\nu},K_{\mu\nu})$ function. 
An analogous situation was found in Ref.~\cite{CPTLTiSh} in the 
Lorentz/CPT violating electrodynamics. In that case the full vector 
field contribution was contained in a single function involving the 
CPT-even violating parameter $k_F^{\mu\nu\al\be}$ and curvature 
tensors. Additionally, in that work it has been shown that the new term 
not affect the dynamics for FLRW metrics with generic spatial 
curvature values $k = 0, \pm 1$. This negative result concerning the 
effect of the new term in the homogeneous and isotropic space-time is expected, 
since the violating fields defines a preferable 
direction in that background. Therefore, this fact can also be seen as 
an additional test for our huge algebraic calculations. Indeed, in our 
case it is not difficult to show that 
$L(g^{FLRW}_{\mu\nu},K_{\mu\nu}) \, = \, 0$. Besides that, one can 
expect that the new terms can cause some modifications in the 
equations for cosmic perturbations during the inflationary epoch and, 
especially, for gravitational waves. The study of gravitational waves 
in the anomaly-induced effective action formalism has been done 
systematically in Ref.~\cite{waveana} and its generalization with the 
presence of the extra Lorentz violating term would be certainly a 
potentially interesting problem. One can expect relevant different 
contributions which can lead to some new constraints on the 
symmetry-breaking parameter $K_{\mu\nu}$. Regardless of the serious 
technical difficulties of this program, it does not look unreliable in 
practice.

\section{Conclusions}
\label{con}

Let us summarize the results obtained. We have calculated the vacuum 
one-loop divergences for the Lorentz/CPT violating scalar field theory 
in curved background. The symmetry-violating parameters were treated as fields, 
rather than constants. The practical calculations have been performed 
for scalars with minimal and nonminimal interaction with gravity by 
application of functional methods and the Schwinger-DeWitt technique 
\cite{dewitt,bavi85}. For minimal real scalars the solution for the 
one-loop counterterms was found in a closed form, while for nonminimal 
complex scalar field the solution has been obtained in the first order 
in the small symmetry-breaking parameters. It turns out that the 
CPT-odd violating field do not contribute for the vacuum divergences 
at that order. All contribution to the renormalization of the vacuum 
comes from the dimensionless parameter $K^{\mu\nu}$, an analogous 
situation to what happens in Lorentz violating QED, where only the 
CPT-even parameter $k_F^{\mu\nu\al\be}$ contributes 
\cite{CPTLGuiSh,CPTLSh,CPTLTiSh}. At same time, we expect that the odd 
parameter becomes relevant in the interacting theory and/or in higher 
orders in Lorentz/CPT symmetry breaking fields. Also, the minimal form 
of the gravitational action requested by the renormalization of 
violating scalar field theory in curved space-time was established 
based on our previously one-loop calculations. In particular, the 
effect of some Lorentz violating gravitational terms which are 
necessary at the quantum level were already discussed in short range 
gravity limit \cite{Bailey:2006fd,HGLCPTNew} and at quantum gravity 
level \cite{Hernaski:2014lga}. At the next stage, a similar analysis 
for the other missing new terms in Eq.~\eq{arel} and those coming 
from the photon sector \cite{CPTLTiSh} would be a very interesting 
exercise, and we hope that with such analysis some new bounds on the 
gravitational Lorentz violating parameters will be established.

The derivation of one-loop divergences for scalar fields with 
nonminimal gravitational coupling also opens the way to study the 
conformal anomaly and anomaly-induced effective action of gravity, 
whose derivations did not bring up serious obstacles. The anomaly 
integration proceeds with minimal changes compared to the known 
procedure, since the new Lorentz violating term is conformal. At the 
one-loop level the anomaly is given by an algebraic sum of the 
contributions of massless conformal invariant fields of spins $0$, 
$1/2$, $1$. The expression obtained, Eq.~\eq{finaction}, represents 
the scalar field contribution for gravitational effective action and 
together with the photon part \cite{CPTLTiSh} must be completed with 
the fermionic contribution, which we are planning to present elsewhere. 
After that, the use of the corresponding gravitational effective action 
in searching for Lorentz violation in the anisotropies of cosmic 
microwave radiation, coming from the cosmic perturbations in the early 
universe, would represent a promising area for application of our 
results. 

\section*{Acknowledgements}
The author wishes to acknowledge 
Coordena\c{c}\~ao de Aperfei\c{c}oamento de Pessoal de N\'ivel Superior
(CAPES) for the support through the PNPD program.
The author
is also grateful to Ilya L. Shapiro and o Breno Loureiro Giacchini 
for useful discussions.

\appendix

\section{Intermediary universal functional traces results}
\label{apA}

To obtain the divergent part of the nonminimal piece of one-loop 
effective action \eq{nmH0}, we shall use the table universal 
functional traces which are an important part of the generalized 
Schwinger-DeWitt technique. The mentioned table correspond to the 
formulas $(4.53)$ up to $(4.61)$ of reference \cite{bavi85} (note that 
here we use opposite sign notations). Using these formulas the 
divergences in each term of Eq.~\eq{nmH0} can be directly calculated. 
After some algebra we obtain
\beq
\n{ti}
\hspace{-0.5cm}
\Tr K^{\mu\nu} \, \na_\mu \na_\nu \,
\frac{1}{\Box} \Big|_{div}
&=&
\frac{2i\mu^{n-4}}{\epsilon}\,
\int d^n x \sqrt{-g} \, \Big\{
K^{\mu\nu}\,
\Big(\frac{1}{90} \, R^{\al\be}\, R_{\al\mu\be\nu}
+\frac{1}{90} \, R_{\al\be\rho\mu} 
\, R^{\al\be\rho}_{\,.\,.\,.\,\nu}
\nonumber
\\
&-&
\frac{1}{45} \, R_{\mu\al} \, R^{\al}_\nu
+ \frac{1}{36} \, R \, R_{\mu\nu}
+ \frac{1}{60}\, \Box R_{\mu\nu}
+ \frac{1}{20}\, \na_\mu \na_\nu R\Big)
\\
&-&
\frac{K}{2} \Big( \frac{1}{180} \, R_{\mu\nu\al\be}^2
- \frac{1}{180} \, R_{\al\be}^2 + \frac{1}{72} \, R^2
+ \frac{1}{30} \, \Box R \Big)\Big\}\,,
\nonumber
\eeq
\beq
\hspace{-0.3cm}
\Tr \left( \xi R - m^2\right)
K^{\mu\nu} \, \na_\mu \na_\nu \,
\frac{1}{\Box^2} \Big|_{div} &=&
\frac{i\mu^{n-4}}{3\epsilon}\,
\int d^n x \sqrt{-g} \, \Big\{
\frac{1}{2}\,K \,\xi\, R^2
- \xi\, K^{\mu\nu}\, R\, R_{\mu\nu}
\nonumber
\\
&+& m^2 \,K^{\mu\nu}\, R_{\mu\nu}
- \frac{m^2}{2} \,K\,R
\Big\}
\,,
\eeq
\beq
&&
\Tr 
\left( m^4 - 2m^2\, \xi\, R + \xi^2 \, R^2 
- \xi \, \Box R \right) K^{\mu\nu} \,
\na_\mu \na_\nu \, \frac{1}{\Box^3}
\Big|_{div} =
\nonumber
\\
&&
\hspace{1cm}
-\frac{i\mu^{n-4}}{2\epsilon}\,
\int d^n x \sqrt{-g} \, \left( m^4 - 2m^2\, \xi\, R + \xi^2 \, R^2 
- \xi \Box R \right) K
\,,
\eeq
\beq
4 \Tr \, \xi \, (\na^\mu \na^\nu R) K^{\al\be}
\, \na_\al \na_\be \na_\mu \na_\nu \, 
\frac{1}{\Box^4} \Big|_{div} &=&
-\frac{i\mu^{n-4}}{3\epsilon} \,\,
\int d^n x \sqrt{-g} \,
\{\xi\, K\, \Box R
\nonumber
\\
&+& 2 \,\xi\, K^{\mu\nu} \, \na_\mu \na_\nu R
\}
\,,
\eeq
\beq
\Tr \, \xi \, K^{\mu\nu} \, (\na_\mu \na_\nu R)
\, \frac{1}{\Box^2}
\Big|_{div} =
-\frac{2i\mu^{n-4}}{\epsilon}\,
\int d^n x \sqrt{-g} \,\,
\xi \, K^{\mu\nu} \, \na_\mu \na_\nu R
\,,
\eeq
\beq
\n{tf}
- 4 \, \Tr \xi \, K^{\mu\al}\, (\na_\al \na^\nu R) 
\, \na_\mu \na_\nu \frac{1}{\Box^3}
\Big|_{div} =
\frac{2i\mu^{n-4}}{\epsilon}\,
\int d^n x \sqrt{-g} \,\,
\xi \, K^{\mu\nu} \, \na_\mu \na_\nu R
\,.
\eeq
By using equations \eq{nmH0} and relations \eq{ti}$-$\eq{tf} one can 
obtain the result \eq{div-nm}.

\section{Proof of conformal invariance of the Lorentz violating 
\texorpdfstring{$L(g_{\mu\nu},K_{\mu\nu})$}{L}-term}
\label{apB}

Let us present here the proof of conformal invariance \eq{conI}. From 
the technical side, this is not a trivial task since the expression \eq{L} is quite complicated, 
then we are going to expose some details 
concerning the needed conformal transformation rules. Since the 
conformal group is a one-parameter Lie group, one can restrict 
our considerations to the infinitesimal version of transformation \eq{tc1}. 
Disregarding the higher orders in $\si$ and superficial terms, after 
some tedious algebra we arrive at the following transformation rules 
for each term present in \eq{L}:
\beq
(\sqrt{-g} \, K^{\mu\nu} R_{\mu \rho \al\be} \, R_{\nu}^{\,.\,\rho\al\be})'
&=& \sqrt{-g} \left[
K^{\mu\nu} R_{\mu \rho \al\be} 
\, R_{\nu}^{\,.\,\rho\al\be}
- 4 K^{\mu\nu} \, R_\nu^\rho \, \na_\mu \na_\rho \si  
\nonumber \right.
\\
&+& \left. 4 K^{\mu\nu} R_{\mu\al\be\nu} \, \na^\al \na^\be \si 
+ \dots \right],
\eeq
\beq
\hspace{-1.0cm}
( \sqrt{-g}\, K^{\mu\nu} R^{\al\be} \, R_{\al\mu\be\nu})' 
&=& \sqrt{-g} \left[ 
K^{\al\be} R^{\mu\nu} \, R_{\al\mu\be\nu} 
+ 2 K^{\mu\nu} R_\nu^\rho \, \na_\mu \na_\rho \si 
- K^{\mu\nu} R \, \na_\mu \na_\nu \si
\right.
\\
&-& \left. K R^{\mu\nu} \, \na_\mu \na_\nu \si 
- K^{\mu\nu} R_{\mu\nu} \, \Box \si 
+ 2 K^{\mu\nu} R_{\mu\al\be\nu} \, \na^\al \na^\be \si 
+ \dots \right],
\nonumber
\eeq
\beq \hspace{-1.0cm}
(\sqrt{-g}\, K^{\mu\nu} R_{\mu\al} \, R^{\al}_\nu)' 
&=& \sqrt{-g} \left[
 \, K^{\mu\nu}\, R_{\mu\al} \, R^{\al}_\nu
- 4 K^{\mu\nu} R_\nu^\rho \, \na_\mu \na_\rho \si 
- 2 K^{\mu\nu} R_{\mu\nu} \, \Box \si
+ \dots \right],
\eeq
\beq \hspace{-0.8cm}
(\sqrt{-g} \, R \, \na_\mu \na_\nu K^{\mu\nu})' 
&=&  \sqrt{-g} \left[
R \, \na_\mu \na_\nu K^{\mu\nu}
- 2 K^{\mu\nu} R \, \na_\mu \na_\nu \si 
- 6 \, \na_\mu \na_\nu K^{\mu\nu} \, \Box \si
\nonumber \right.
\\
&-& \left. 6 \, K^{\mu\nu} \na_\mu R \,  \na_\nu \si 
+ K \, \na_\rho R \, \na^\rho  \si 
+ \dots \right],
\eeq
\beq \hspace{-0.5cm}
(\sqrt{-g}\, R_{\mu\nu} \, \Box K^{\mu\nu})' 
&=& \sqrt{-g} \,[
 R_{\mu\nu} \, \Box K^{\mu\nu} 
- 4 K^{\mu\nu} R_\nu^\rho \, \na_\mu \na_\rho \si 
- 2 K^{\mu\nu} R_{\mu\nu} \, \Box \si
\nonumber
\\
&-& 4  K^{\mu\nu} R_{\mu\al\be\nu} \, \na^\al \na^\be \si 
- 2 \,  \na_\mu \na_\nu K^{\mu\nu} \Box \si
- 2 \, K^{\mu\nu} \na_\mu R \,  \na_\nu \si 
\\
&-& K \, \Box^2 \si 
+ \dots ] \, \nonumber
\eeq
and
\beq
&& \hspace{-0.2cm}
\Big[- \sqrt{-g} \, \frac{K}{2}\,\Big(\frac{1}{180}\, R_{\mu\nu\al\be}^2 
- \frac{1}{180}\,R_{\mu\nu}^2
+ \frac{1}{180}\, \Box R \,
\Big) \,\Big]' =
\sqrt{-g}\, \Big[-\frac{K}{2}\,
\Big(\frac{1}{180}\, R_{\mu\nu\al\be}^2 
\nonumber
\\
&&
\hspace{0.07cm}
-\, \frac{1}{180}\,R_{\mu\nu}^2
+ \frac{1}{180}\, \Box R \,
\Big)
+ \frac{1}{90}\, K \,R^{\mu\nu} \,\na_\mu \na_\nu \si
- \frac{1}{180}\, K \, \na_\rho R \, \na^\rho \si
+ \frac{1}{60}\, K \, \Box^2 \si
\nonumber \\
&&
\hspace{0.07cm}
+ \dots \,\Big] .
\eeq
Substituting the above formulas into \eq{L}, we find the conformal 
invariance \eq{conI}.


\end{document}